\documentclass[aps,pra,twocolumn,groupedaddress]{revtex4}

\usepackage{graphicx}
\usepackage{dcolumn}
\usepackage{bm}

\begin{document}
\title{Integrating cavity quantum electrodynamics and ultracold-atom chips with on-chip dielectric mirrors and temperature stabilization}
\author{T.P. Purdy}
\email{tpp@berkeley.edu}
\author{D.M. Stamper-Kurn}
\affiliation{Department of Physics, University of California, Berkeley CA  94720}
\date{July 1, 2007}

\begin{abstract}
We have fabricated an atom chip device which combines the
circuitry for magnetic trapping of cold atoms with high-finesse optical resonators suitable for cavity QED in the single-atom strong coupling regime.  Fabry-Perot optical resonators with finesse
$\mathcal{F} \geq 2 \times 10^5$ were formed between a
micropatterned on-chip planar mirror with lateral dimension of
$\leq 100\,\mu$m and a curved mirror suspended above the chip.  The strong and rapid thermal coupling between on-chip electrical and optical elements was utilized to stabilize the cavity mirror separation with servo bandwidth exceeding 100 kHz during simulated operation of the atom chip.


\end{abstract}
\maketitle

Atom chips \cite{drnd98,reic99mag,folm00,ott01chip,hans01}, which incorporate microfabricated electromagnets, electrodes and magnetic materials to trap atoms in vacuum above the chip surface, stand to expand the scientific reach of ultracold atomic gases. Given that optical manipulation and detection is essential to many ultracold atom experiments, the utility of atom chips is enhanced by the addition of optical elements onto the chips.

In particular, on-chip optical resonators would allow for highly sensitive, high bandwidth, and localized detection of atoms, improving the precision and allowing explorations of quantum-limited measurements.  Low-finesse cavities can improve detection sensitivity \cite{stei06} even to the single atom level \cite{hora03chip,long03chip,lev04nano,tepe06res}, although such detection is destructive in that detected atoms are perforce heated or changed in their internal state.  In contrast, higher finesse cavities with small mode size enable \emph{nondestructive} atom detection at the single atom level, provided that the single-atom strong coupling regime ($C \gg 1$) is attained.   Here, the single-atom cooperativity is defined as $C=g^2/ 2 \kappa \Gamma$, where $g$ is the vacuum Rabi frequency, and $\kappa$ and $\Gamma$ are the cavity and atomic coherence decay rates, respectively. Indeed, in this regime, myriad protocols based on cavity quantum electrodynamics (CQED) allow an interface between material and optical representations of quantum information.

\begin{figure}
\begin{center}
\includegraphics[width=8.5cm]{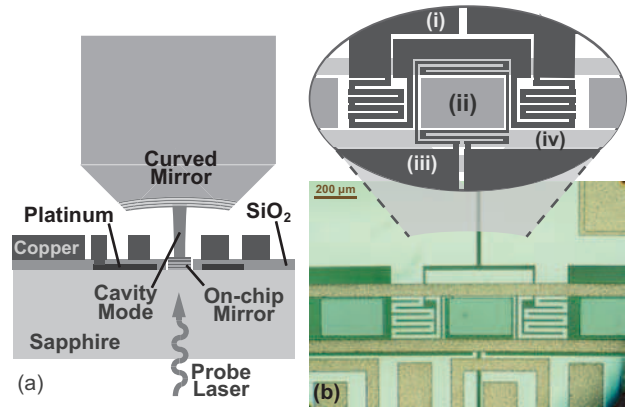}\\
\caption{(a) A CQED/atom chip includes electromagnet wires and a
small planar mirror microfabricated on a sapphire substrate.  A
curved mirror suspended above the chip completes a vertically
oriented Fabry-Perot resonator in the strong-coupling regime of
CQED.  (b) Microscope image of finished device
with inset schematic showing portions of the platinum layer obscured
in the photograph. (i) thermometer wire, (ii) dielectric mirror pad, (iii) heater wire, (iv) waveguide wire} \label{cavity_schematic}
\end{center}
\end{figure}

Several approaches have been pursued for achieving the strong-coupling regime of CQED in a manner compatible with magnetic or optical micromanipulation of ultracold atoms.  Fabry-Perot resonators have been have been employed in atom chip experiments using both conventional macroscopic \cite{tepe06res} and novel microscopic \cite{stei06} mirrors, attaining cooperativity as high as $C = 2.1$. Other microscopic Fabry-Perot resonators with higher cooperativity have also been constructed \cite{trup05apl,treu06fort}. Single atoms have been detected using monolithic microtoroid resonators \cite{aoki06}, and resonators employing artificial photonic bandgap materials \cite{vuck01} have been fabricated; both achieve the strong coupling regime and, in principle, are integrable with atom chip elements.

Here, we describe a microfabricated CQED/atom chip combining
magnetic atom traps and a high-finesse Fabry-Perot cavity, with a single atom cooperativity of up to $C = 50$ for the $^{87}$Rb D2 transition at a wavelength $\lambda\!=\!780$ nm.  The cavity utilizes an on-chip planar mirror, patterned with lateral dimensions of $100 \, \mu$m or below.  The second, curved mirror that completes the cavity is mounted above the chip with its optical axis perpendicular to the chip surface (Fig.\
\ref{cavity_schematic}). Conductors are fabricated on the mirror substrate, creating a two-wire magnetic waveguide and magnetic conveyor belt \cite{hans01conveyor} straddling the on-chip mirror.  Ohmic heating in these electromagnet wires leads to substantial displacement of the on-chip mirror.  However, we turn this strong thermal coupling to an advantage as a means of deliberately actuating the cavity mirror via an on-chip temperature stabilization circuit.  The cavity resonance is thereby stabilized with a servo bandwidth exceeding 100 kHz.  Our methods for microfabricating and rapidly actuating high reflectivity optics represent an advance in the miniaturization of optical components, with applications beyond ultracold atom experiments.

In designing the on-chip mirror, one encounters a coincidence of length scales.  On one hand, the typical $\sim 5$ cm radius of curvature for the curved mirror sets a waist of $w_0 \sim 20 \, \mu$m for the optical mode supported by a cavity with mirror spacing in the 10 -- 100 $\mu$m range.  In order to maintain a cavity finesse of $\mathcal{F}\geq 10^5$, a clear aperture of radius
$\simeq 50 \,\mu$m is then required to limit diffraction losses at
the aperture edge. On the other hand,  confining rubidium atoms
magnetically to linear dimensions $d \ll \lambda$, so that the
coupling strength between the atom and the standing-wave cavity mode
is well determined, requires that electromagnets be placed within $100 \, \mu$m of the atoms. Here, we assume the electromagnets operate at current densities of  $\sim 10^6 \, \mbox{A/cm}^{2}$,  a practical limit for atom chip wires
\cite{fort07rmp}.  This consideration sets a maximum size for the
clear aperture of the on-chip mirror.

The CQED/atom chip fabrication starts with a two inch diameter, 4 mm
thick, $c$-axis cut sapphire wafer, which is superpolished on one
surface and coated with a high reflectivity multilayer dielectric
mirror coating consisting of alternating layers of SiO$_2$ and
Ta$_2$O$_5$ (Research Electro-Optics, Boulder, CO).  This mirror
coating, with total thickness of over 5 $\mu$m, was optimized for
highest reflectivity at 780 nm, and some samples were measured to
have total scattering, absorption, and transmission losses below 10 ppm per reflection.

Small on-chip mirrors were formed by etching away the
high-reflectivity dielectric coating except in selected areas that
form the remaining ``mirror pads.''  To define these pads and
protect their surfaces from subsequent processing, a 400 nm layer of
high purity aluminum was first thermally evaporated onto the mirror
surface. This layer was then patterned, and the exposed dielectric
coating was removed in a reactive ion, parallel plate plasma etcher
operating with 100 sccm CF$_4$ and 10-20 sccm O$_2$, at a pressure
of about 85 mTorr and an RF power density of about 0.4 W/cm$^2$, and
with the chip at a temperature of 120$^\circ$C.  Since the etch rate
for sapphire was negligible compared to that for the dielectric
layers (80 nm/min), such etching left exposed the flat substrate in
all unprotected regions of the mirror.  Lateral etching of about 5
$\mu$m on the margins of the mirror pads was observed.

\begin{figure}
\begin{center}
\includegraphics[width=8cm]{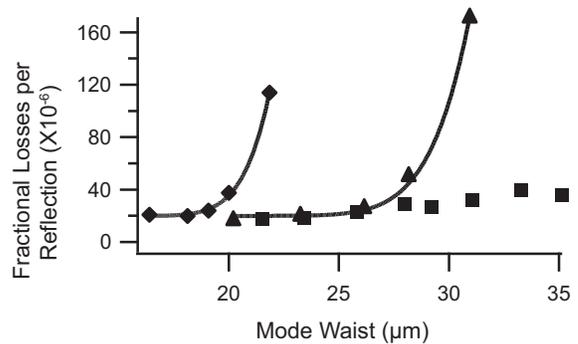}\\
\caption{Losses in microfabricated mirrors with nominal radii of 50 (diamonds), 70 (triangles), and 100 $\mu$m (squares) were measured for variable cavity mode waists.  The 20 ppm total losses per reflection from the on-chip mirror observed at the smallest cavity spacing is consistent with that observed for the mirror prior to microfabrication.  Increased diffraction losses at larger cavity spacing, with lines showing fits to a simple model, indicate effective mirror radii of 47 and 65 $\mu$m for the two smallest mirrors, respectively, consistent with the observed lateral etching of the mirror pads during their fabrication.}
 \label{fin_plot}
\end{center}
\end{figure}

To test the mirrors microfabricated in this manner, we patterned a sapphire wafer with mirror pads of dimensions ranging between 40 and 250 $\mu$m.  The aluminum mask was then removed using a commercial etchant. Fabry-Perot cavities were formed with each on-chip mirror using a 5 cm radius of curvature mirror positioned above the chip surface. The cavity finesse was then measured for varying cavity lengths by probing with a grating-stabilized diode laser at wavelengths near $\lambda = 780$ nm . At the smallest cavity spacing, for which the supported cavity mode had the narrowest waist, these cavities attained a finesse of $\mathcal{F} \geq 2 \times 10^5$, equal to that attained prior to microfabrication given the parameters top curved mirror.  The cooperativity of such cavities is about 50 for a 25 $\mu$m mirror spacing.  Here the finesse is limited mainly by scattering and absorption in the on-chip dielectric coating.  We have, however, obtained sapphire wafers with  higher quality mirror coatings.  If such a wafer was employed as a CQED/atom chip, the finesse and cooperativity of the on-chip cavity would be improved by at least a factor of two.

As the spacing between the cavity mirrors was increased, the transverse waist of the supported cavity mode increased, and the cavity finesse was diminished by diffraction losses off the edges of the on-chip mirror (Fig.\ \ref{fin_plot}).  We quantify these losses per reflection as the fractional intensity of a Gaussian beam with $1/e^2$ radius $w_0$ that falls outside the effective radius $a$ of the on-chip mirror. The total round-trip loss $\delta_c = 2 \pi / \mathcal{F}$ is then given as $\delta_c = e^{-2 a^2 / w_0^2} + \delta_1+\delta_2$ where $\delta_1$, $\delta_2$ are the remaining, spacing-independent loss of each of the two cavity mirrors.  By fitting to this expression for $\delta_c$, we obtained experimentally the usable radius of each on-chip mirror. This was found to be 3 -- 5 $\mu$m smaller than the nominal radius of the microfabricated mirror, consistent with the observed lateral etching of the mirrors during their microprocessing.

The birefringence of the mirror coating was  monitored before and
after processing.  Circular mirror pads exhibited linear
birefringent phase shifts of $\simeq 10^{-5}$ rad per reflection,
similar to those observed on the unprocessed mirrors.  For mirror
pads fabricated with a rectangular shape, the linear birefringence
was increased to $\simeq 10^{-4}$ rad, with principal axes
correlated with the orientation of the rectangle.  This effect was
observed for mirror pads formed both on sapphire and on glass
substrates, and is ascribed to strains induced in the mirror
coatings by the asymmetric mirror shape.

In designing a complete CQED/atom chip incorporating such mirror pads the inevitable thermal coupling of the mirrors to the current-carrying, heat-generating wires on the chip must be considered.  To quantify this coupling we calculate the expected temperature variation  $T(r, \omega) \, e^{-\imath \omega t}$ due to an AC heat source on the surface at a distance $r$
and varying at angular frequency $\omega$. For a point heat source
generating power $P_0 \, e^{- \imath \omega t}$ on the surface of a
half space of material, we obtain as a solution to the heat equation
\begin{equation}
T(r,\omega) = \frac{P_0}{2 \pi k} \frac{e^{- \sqrt{\frac{i \omega r^2}{\alpha}}}}{r}
\end{equation}
while for a line source generating power per length of $A_0 \, e^{-
\imath \omega t}$ we find
\begin{equation}
 T(\rho,\omega) = \frac{-i A_0}{2 k} \left[ J_0\left(\sqrt{\frac{-i \omega \rho^2}{\alpha}}\right) -i Y_0\left(\sqrt{\frac{-i \omega \rho^2}{\alpha}}\right)
 \right]
\end{equation}
Here $J_0$ and $Y_0$ are Bessel functions of the first and second
kind respectively, $\rho$ is the distance to the line, $k = 40 \, \mbox{W}/\mbox{m} \mbox{K}$ is the thermal conductivity of sapphire, and $\alpha = 1.3 \times 10^{-5}\,
\mbox{m}^2/\mbox{s}$ is its thermal diffusivity.

In the limit $\omega\!\rightarrow\!0$, these expressions indicate the steady-state temperature increase at the chip surface during its operation.  For example, the DC temperature rise at a mirror pad located 100 $\mu$m from copper wires running 3 A through a 300 $\mu$m$^2$ wire cross section is estimated as 20 K. Under these conditions, the expansion of the substrate below the mirror would lift the mirror surface by about 125 nm.  This displacement is enormous compared to the mirror displacement of $\lambda / 2 \mathcal{F} \simeq 4$ pm  required to shift the cavity resonance by one cavity linewidth.  For $\omega \gg  \alpha / r^2 \simeq 2 \pi \times 200$ s$^{-1}$ the temperature response for both geometries is exponentially suppressed.  For the example mentioned above this implies a rapid timescale $\tau=r^2/\alpha \simeq 1$ ms for the expansion after a sudden switch-on of the atom chip electromagnets. Altogether, the slew rate of $\simeq 100$ nm/ms that would be required to actively stabilize the cavity resonance during operation of the atom chip presents a formidable technical challenge.

To address this  challenge, elements for a high-speed temperature
stabilization circuit were fabricated as part of the CQED/atom chip.
Platinum wire heaters and Resistance Temperature Detectors (RTD)
were placed directly onto the sapphire substrate, near the mirror, to
monitor and control the substrate surface temperature.

The complete CQED/atom chip device (Fig.\ \ref{cavity_schematic})
was fabricated in two steps. First, microfabricated mirrors were
created using the process described above.  Then, with mirror pads
still protected by aluminum, standard atom-chip fabrication methods
were employed \cite{lev03chip} to produce the heater, thermometer,
and magnet wires on the sapphire substrate left bare by the previous
plasma etch.

The platinum RTD and heater wires, 100 nm thick, were patterned using electron beam evaporation and a lift-off process. This method was found to be suitable for patterning 5 $\mu$m wide features within 10 $\mu$m of the mirror locations, despite steps of almost 6 $\mu$m in height from the sapphire substrate to the to surface of
the mirror.  Resistances of several k$\Omega$ were obtained in an area as small as .04mm$^2$.  A 1$\mu$m SiO$_2$ was next applied to the wafer
via plasma enhanced chemical vapor deposition, which requried the
entire wafer to reach a temperature of 350$^\circ$C.  This layer
insulated the platinum wires from overlaid conducting layer and
also partially insulated the electromagnet wires from the substrate,
slowing the thermal coupling to the on-chip mirrors. Holes were patterned through the SiO$_2$ layer to make electrical contacts between the platinum wires and overlaid copper leads.  Copper magnet wires were
electroplated to a thickness of 5 -- 10 $\mu$m through a photoresist
mold.  Finally the mirror surfaces were uncovered via a wet etch,
and the finished wafer was cleaned in acetone, isopropyl alcohol,
and deionized water.

The operation of this CQED/atom chip was tested under conditions
simulating the operation of the chip in cold-atom experiment.  A
Fabry-Perot cavity was constructed as before, employing the on-chip
mirror surrounded by platinum and copper wires.   The whole assembly
was mounted on a vibration isolated copper heatsink, and
measurements were performed in open air.

The optical cavity was stabilized by monitoring the transmission of
a probe laser and feeding back to either a piezoelectric transducer
(PZT) that displaced the top mirror or the platinum heater that
displaced the on-chip mirror.  Following the example of other CQED
experiments \cite{ye99trap,fisc02}, the cavity was probed with laser
light far detuned from the atomic transitions of rubidium.  At this
laser wavelength of 850 nm, the cavity linewidth was about 10 times
larger than at 780 nm.  Thus we required that the cavity resonance
be maintained to just a few percent of its linewidth at 850 nm.

\begin{figure}
\begin{center}
\includegraphics[angle=90,width=8.cm]{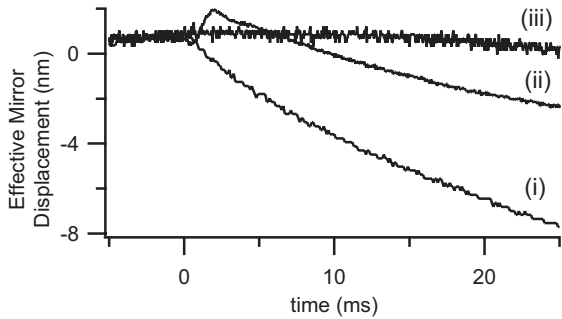}\\
\caption{Response of cavity resonance to an 0.5 A current pulse
through the waveguide wires.  The PZT voltage necessary to stabilize
the cavity resonance, indicative of thermally induced displacements
of the on-chip mirror, is recorded when the heater wire is either
(i) not used, (ii) used to feedback to temperature variations
measured at the RTD, or (iii) used in a feed-forward scheme.}
 \label{thermometer_plot}
\end{center}
\end{figure}

We tested three different schemes for stabilizing the cavity during
rapid variations of currents in the magnetic waveguide wires.  In
the first scheme, we stabilized the temperature, as measured by the
RTD, using the heater wire, while independently stabilizing the
cavity transmission  using PZT actuation of the curved mirror. While
the temperature stabilization operated with servo bandwidth of
several kHz, we found that the response of the \emph{cavity
resonance} to varying thermal conditions was only moderately
diminished by this stabilization (Fig.\ \ref{thermometer_plot}).

A second stabilization scheme added a feed-forward component to the
method described above.  Here, a signal proportional to the magnet
wire current was filtered, inverted, offset, and fed into the heater
wire.  This feed-forward scheme was tuned so as to suppress the
response of the cavity to rapid thermal changes by over an order of
magnitude for a short time after the magnet wire currents were
switched.

The most successful stabilization scheme employed a combination of
both PZT and thermal actuation of the two cavity mirrors to
stabilize directly the optical cavity resonance. The response of the
cavity resonance position to actuation by the on-chip heater showed
the expected roll off in frequency starting at 200Hz, but maintained a smooth
feedback phase beyond 200 kHz, showing no discernible coupling to
mechanical resonances. Remarkably, we were thus able to apply
straightforward thermal feedback with servo bandwidth of over 100
kHz.

\begin{figure}
\begin{center}
\includegraphics[width=8.cm]{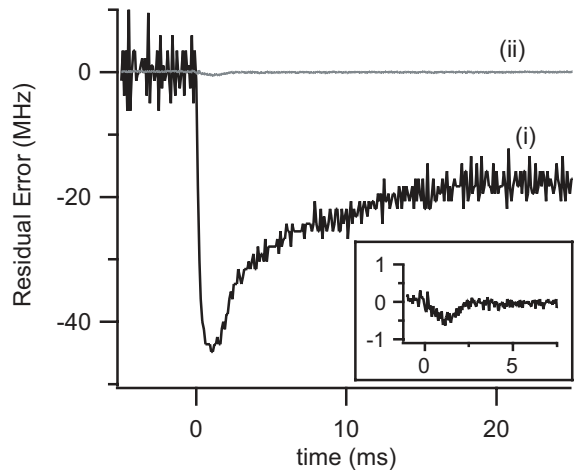}\\
\caption{Deviation of cavity resonance frequency due to
current pulses through the waveguide electromagnets.
(i) Using only feedback to the PZT actuation of the curved cavity mirror, a sudden 0.5 A current pulse displaced the cavity resonance by more than the 35 MHz cavity linewidth for over 100 ms.  (ii) With feedback employing both PZT and fast thermal actuation of the cavity mirrors, the cavity resonance remained stablized well within even the few MHz linewidth of a very high finesse mode for sudden pulses of 3 A through the electromagnet wires. Inset shows the small residual disturbance from the 3 A pulse} \label{err_response}
\end{center}
\end{figure}

As shown in Fig.\ \ref{err_response}, this last stabilization scheme
strongly suppressed both rapid transients and slow drift of the
cavity resonance frequency (i.e.\ the effective cavity spacing) in
response to the switching heat loads of an operational atom chip.
Over 3 A of current within 10's of microns of the on-chip mirror,
generating $\sim$0.3 W/mm of heat, were toggled with little effect
on the measured cavity transmission of a fixed-frequency optical
probe.  In contrast, the other stabilization schemes  required that
currents be ramped up over 100 ms or longer for the cavity to remain
stabilized.  At the center of the cavity mode the estimated magnetic field gradient in a direction transverse to the waveguide axis for 3 A of current through the guide wires is about 4000 G/cm.  We believe this will be sufficient to localize cold atoms to a dimension much less than $\lambda$.

In utilizing this thermal feedback scheme in a cold-atom experiment, attention must be paid to the varying magnetic fields produced by the on-chip heating element that would perturb the magnetically trapped atoms. While in our prototype, the heater generated 40 mG/mA at a 50$\mu$m distance, improved designs would reduce this figure to below 1 mG/mA. Moreover, heater wires can be operated with an alternating current at a frequency far from the vibration frequencies of trapped atoms, and also from Larmor precession frequency at the bias field of the magnetic trap.  The net effect of the residual oscillating magnetic field is to create adiabatic potentials \cite{zoba01,schu05nphys} at sub-nanokelvin levels.

In conclusion, we have demonstrated the fabrication of magnetic
trapping circuitry to be compatible with the direct integration of
micron-scale high reflectivity mirrors.  The thermal coupling
between the atom trapping and optical elements has been accounted
for and utilized to rapidly actuate an optical cavity.  The scalable
microfabrication processes used to create this CQED/atom chip allow
for many cavities to be integrated onto a single atom chip,
enhancing the potential of quantum optical and quantum atom optical
devices.  Furthermore, the rapid thermal actuation of low-loss optical components demonstrated in this work may find use in applications not related to atom chip experiments, e.g.\ those requiring high-finesse resonators to be stabilized even in noisy environments.

We would like to thank Daniel Brooks for assistance in this experiment.  This work was
supported by the AFOSR under Grant No.\ FA9550-04-1-0461, and by the
David and Lucile Packard Foundation.



\end{document}